\documentstyle[aps,prc,amssymb,floats,epsfig]{revtex}

\begin{document} 

\def\dalphg{$d(\alpha,\gamma)^6{\rm Li}$}
\def\tag{$^3{\rm H}(\alpha,\gamma)^7{\rm Li}$}
\def\hag{$^3{\rm He}(\alpha,\gamma)^7{\rm Be}$}
\def\htwo{$^2{\rm H} $}
\def\h3{$^3{\rm H} $}
\def\he3{$^3{\rm He} $}
\def\li7{$^7{\rm Li} $}
\def\be7{$^7{\rm Be} $}
\def\b8{$^8{\rm B} $}

\draft
\title{Radiative alpha-capture cross sections from realistic nucleon-nucleon 
interactions and variational Monte Carlo wave functions}
\author{Kenneth M. Nollett\footnote{Present address:  MC 130-33 California
Institute of Technology, Pasadena, CA~~91125}}
\address{Department of Physics,Enrico Fermi Institute, 
	 The University of Chicago, Chicago, IL~~60637-1433\\ 
	 and\\ 
	 Physics Division, Argonne National Laboratory,
	 Argonne, Illinois 60439}
\date{\today}
\maketitle

\begin{abstract}
We report the first calculations of cross sections for the radiative
capture reactions \tag\ and \hag\ below 2 MeV which use wave functions
derived from realistic nucleon-nucleon interactions by the variational
Monte Carlo technique.  After examining several small corrections to
the dominant $E1$ operator, we find energy dependences for the
low-energy $S$-factors which agree reasonably with experimental
measurements.  There is no contradiction with the previous theoretical
understanding of these processes, but the zero-energy derivative of
the \tag\ $S$-factor is smaller than in most models.  While this
method can in principle predict cross section normalizations, the
normalizations of our results are mostly too low.
\end{abstract}

\section{Introduction}

Electromagnetic captures of alpha particles on \h3 and \he3 are
important processes in astrophysics.  Together, they are responsible
for all $^7$Li production in the standard big bang nucleosynthesis
(BBN) calculation.  Because their cross sections are also difficult to
measure in the laboratory at the relevant energies (20--500 keV), they
are the major sources of uncertainty in the calculated primordial
$^7$Li abundance \cite{nollettburles,bntt}.  \hag\ is also important
for predicting the production rate of \b8 and $^7$Be neutrinos in the
sun.  Accurate knowledge of its rate at solar energies ($\sim 20$ keV)
is therefore important for studies of the solar neutrino problems
\cite{adelberger}.

Recent theoretical studies of these cross sections have used two
closely-related approaches.  Potential
models~\cite{dubovichenko95,kim81,buck85,buck88,mohr93} treat alpha
particles and tritons (or \he3 nuclei -- the isospin symmetry of these
systems makes the two processes almost identical) as point particles
and the final states as bound states of these point particles.
Authors of such models generate wave functions from potentials that
fit experimental scattering phase shifts and bound state properties
(binding energies, electromagnetic moments, etc.), and then compute
the cross section as a direct capture.  The resonating-group method
(RGM)
~\cite{kim81,walliser83,kajino84,walliser84,liu81,fujiwara83,kajino86,mertelmeier86,altmeyer88,csoto00}
is fully microscopic, in that it solves an explicitly seven-body
problem with a nucleon-nucleon potential, the parameters of which are
adjusted to reproduce bound-state and resonance properties for the
particular problem being solved.  The name is derived from the choice
of basis states for solving the Schr\"odinger equation, which consists
of one or more partitions of the particles into clusters with internal
harmonic oscillator structure.  The potential models now have a
well-founded justification in the resonating-group work, in the form
of the microscopic potential model~\cite{friedrich78,langanke86}.
Although the RGM models have apparently been very successful in
describing these reactions (the calculation of Kajino \cite{kajino86}
correctly predicted both energy dependence and normalization for the
experiment of Brune et al. \cite{tag-brune}), it is not clear that
this is the final word, because the agreement with the data seems to
be spoiled when the model space is expanded \cite{altmeyer88,csoto00}.

It was shown in the early work of Christy and Duck~\cite{christyduck}
that low-energy radiative captures on light nuclei may be treated to
good approximation as external direct captures, that is, as one-step
processes in which most of the matrix element arises outside the
nuclear interaction radius.  It remains true in more detailed models
that the largest contributions to the matrix elements arise in regions
well outside the range of the internuclear forces.  In principle,
then, the cross section energy dependence is given by convolutions of
(positive- and negative-energy) Coulomb wave functions with the
current operator, while the normalization of the cross section is
determined by the asymptotic normalization of the bound state in the
appropriate clusterization channel.

However, fine details are obviously missing from the external direct
capture model, motivating application of more detailed models that can
predict both energy dependence (which is relatively easy to measure)
and normalization (which is not easy to measure, and which must be
inserted by hand in an external direct capture calculation after
determination by some other means).  Small effects involving the
short-range ($< 5$ fm) behavior of the nuclear wave functions, both in
the alpha-trinucleon channel and in other channels, can affect the
cross sections by several percent.  Such effects are probably behind
the differences between models pointed out in Ref. \cite{csoto00}.  In
principle, there are also small corrections to the leading-order
current operators.  Because of the astrophysical importance of
determining these cross sections, and especially the need for
low-energy extrapolation of \hag\ for solar physics, it is important
to apply new approaches to this problem as they become available, and
compare the results with past efforts.

In this context, recent developments in the physics of light nuclei
are particularly interesting.  There now exist ``realistic''
nucleon-nucleon potentials which describe the $np$ and $pp$ scattering
data, as well as the deuteron, with high precision ({\it e.g.}
Ref.~\cite{wss95}).  Further interactions not describable by two-body
potentials are described by three-nucleon potentials, which have been
adjusted to reproduce residual effects in the energy spectrum of light
nuclei and properties of nuclear matter~\cite{ppcw95}.  Wave functions
have been developed for these potentials in systems with up to eight
nucleons~\cite{a=8}.  This provides an opportunity to approach the
problem of radiative captures on light nuclei using realistic
potentials and the computational techniques that have been developed
to utilize them.

Conversely, astrophysical interest in these processes has resulted in
relatively precise measurements, which make cross section calculations
useful tests of the wave functions in addition to the reproduction of
the static moments, electron scattering properties, and energy spectra
to which they have been compared in the past.  We have already
reported the application of these wave functions to a radiative
capture calculation in a paper on the process \dalphg ~\cite{nws}.

The remainder of this paper describes cross section calculations for
the reactions \tag\ and \hag, using bound-state wave functions derived
from realistic potentials by the variational Monte Carlo method.  It
is organized as follows: In Section \ref{sec:wf}, we describe the wave
functions used to compute the cross sections.  In Section
\ref{sec:operators}, we describe the electromagnetic current operators
and the methods used to compute their matrix elements.  In Section
\ref{sec:results}, we describe the results for cross sections and
branching ratios.  In Section \ref{sec:conclusion}, we examine the
implications of our results.

\section{Wave functions}
\label{sec:wf}

\subsection{Bound states}

We used ground states of \h3, \he3, $^4{\rm He}$, $^7$Li, and $^7$Be
which were found by the variational Monte Carlo (VMC) technique for
the Argonne $v_{18}$ two-nucleon potential (hereafter AV18)
\cite{wss95} and the Urbana IX three-nucleon potential (UIX)
\cite{ppcw95}.  The radiative captures can go to either the ground
state or the (bound) first excited state in both $^7$Li and $^7$Be, so
the first excited states of these nuclei were also needed.  These wave
functions were generated by the same VMC method as the ground states.
The bound-state wave functions have been reported in Refs. \cite{W91}
(triton and $^4$He) and \cite{ppcpw97} (modified here as in
Ref. \cite{nws} to obtain $^7$Li and $^7$Be bound states with desired
asymptotic properties).

The VMC method proceeds by constructing wave functions as products of
pair and triplet correlations between nucleons, and adjusting free
parameters in these correlations to minimize energy expectation values
which are computed by a Monte Carlo integration.  The bound state wave
functions are built from central and operator correlations between
nucleons, acting on a Jastrow wave function,
\begin{equation}
     |\Psi_T\rangle = \left[ 1 + \sum_{i<j<k} \tilde{U}^{TNI}_{ijk} \right]
              \left[ {\cal S}\prod_{i<j}(1+U_{ij}) \right] |\Psi_J\rangle \ ,
\label{eqn:trial}
\end{equation}
where $U_{ij}$ and $\tilde{U}^{TNI}_{ijk}$ are two- and three-body
correlation operators that include spin and isospin dependence and
${\cal S}$ is a symmetrization operator, needed because the $U_{ij}$
do not commute.  The sums and products throughout this Section are
over all nucleons.  For $^4$He, the Jastrow part takes a relatively
simple form:
\begin{equation}
  |\Psi_J\rangle = \prod_{i<j<k \leq 4}f_{ijk}
     \prod_{i<j \leq 4}f(r_{ij}) |\Phi_{\alpha}(0 0 0 0)_{1234}\rangle \ ,
\label{eqn:alpha}
\end{equation}
where $f(r_{ij})$ and $f_{ijk}$ are pair and triplet functions of
relative position only, and $\Phi_{\alpha}(0 0 0 0)$ is a determinant
in the spin-isospin space of the four particles.  Jastrow wave
functions for \he3 and for the triton are constructed analogously, but
the parameters of the $f(r_{ij})$ for these nuclei have been chosen to
minimize energy expectation values of the three-body nuclei rather
than of the alpha particle.  The triton and $^3$He are identical in
our calculation, except for their isospin vectors.  In cases where the
distinction is unimportant, we refer to both nuclei as ``the
trinucleon,'' and denote them both by $\tau$ in subscripts that label
clusters.

For larger nuclei, spatial dependences must be introduced to place
some particles in the p-shell.  The $A=7$ Jastrow wave function is
constructed from scalar correlations multiplying a shell model wave
function,

\begin{eqnarray}
  |\Psi_J\rangle &=& {\cal A} \left\{
     \prod_{i<j<k \leq 4}f^{sss}_{ijk}\ 
     \prod_{n \leq 4}\ \prod_{5\leq m  < 7}\ 
	\prod_{ m < p \leq 7}\ f^{spp}_{nmp}
     \prod_{i<j \leq 4}f_{ss}(r_{ij})\right.  
     \nonumber\\
  &&    \prod_{k \leq 4}\ \prod_{5\leq n \leq 7}\ f_{sp}(r_{kn}) 
	f_{pp}(r_{56})f_{pp}(r_{57})f_{pp}(r_{67})
     \nonumber\\
  && \left.  \sum_{LS[n]} \Big( \beta_{LS[n]} 
     |\Phi_7(LS[n]JMTT_{3})_{1234:567}\rangle \Big) \right\} \ ,
\label{eqn:jastrow}
\end{eqnarray}
where ${\cal A}$ is an antisymmetrization operator over all partitions
of the seven particles into groups of four and three.  For the central
pair and triplet correlations, $f_{xy}(r_{ij})$ and $f^{xyz}_{ijk}$,
the $xyz$ denote whether the particles are in the s- or p-shell.  The
shell model wave function $|\Phi_7(LS[n]JMTT_{3})\rangle$ has orbital
angular momentum $L$, spin $S$, and spatial symmetry $[n]$ 
coupled to total angular momentum
$J$, projection $M$, isospin $T$, and charge state $T_{3}$, and is
explicitly written as
\begin{eqnarray}
|\Phi_{7}(LSJM[n]TT_{3})_{1234:567}\rangle & = &
     |\Phi_{\alpha}(0 0 0 0)_{1234}
     \phi^{LS[n]}_{p}(R_{\alpha 5}) \phi^{LS[n]}_{p}(R_{\alpha 6}) 
	 \phi^{LS[n]}_{p}(R_{\alpha 7})
     \nonumber \\
 &&  \left\{ [Y_{1m_l}(\Omega_{\alpha 5}) Y_{1m_l'}(\Omega_{\alpha 6})
	Y_{1m_l''}(\Omega_{\alpha 7})]_{LM_L} \right.
     \nonumber \\
 &&   \times [\chi_{5}(\frac{1}{2}m_s) \chi_{6}(\frac{1}{2}m_s')
	\chi_{7}(\frac{1}{2}m_s'')]_{SM_S}
     \}_{JM} \nonumber \\
 &&  \times [\nu_{5}(\frac{1}{2}t_3) \nu_{6}(\frac{1}{2}t_3')
	\nu_{7}(\frac{1}{2}t_3'')]_{TT_3}\rangle \ .
\end{eqnarray}
The $Y_{LM}(\Omega)$ are spherical harmonics, $\chi(s,m_s)$ are
spinors, and $\nu(t,t_3)$ are spinors in isospin, while brackets with
subscripts denote angular momentum and isospin coupling.

Particles 1--4 are placed in the s-shell core with only spin-isospin
degrees of freedom, while particles 5--7 are placed in p-wave orbitals
$\phi^{LS[n]}_{p}(R_{\alpha k})$ that are functions of the distance
between the center of mass of the core and particle $k$.  Different
amplitudes $\beta_{LS[n]}$ in Eq. (\ref{eqn:jastrow}) are mixed to
obtain an optimal wave function; for the $J^\pi,T=3/2^-,1/2$ ground
state of $^7$Be, the p-shell can have $\beta_{1\frac{1}{2}[3]}$,
$\beta_{1\frac{1}{2}[21]}$, $\beta_{1\frac{3}{2}[21]}$,
$\beta_{2\frac{3}{2}[21]}$, and $\beta_{2\frac{1}{2}[21]}$ terms.  By
far the largest contribution from these terms, as expected and as
derived by diagonalization of the variational wave functions, is
$\beta_{1\frac{1}{2}[3]}$.  This is true of all the mass-seven,
$T=1/2$ bound states, which are the final states of the radiative
captures in question.

The two-body correlation operator $U_{ij}$ is defined as:
\begin{equation}
     U_{ij} = \sum_{p=2,6} \left[ \prod_{k\not=i,j}f^p_{ijk}({\bf r}_{ik}
              ,{\bf r}_{jk}) \right] u_p(r_{ij}) O^p_{ij} \ ,
\end{equation}
where the $O^{p=2,6}_{ij}$ = ${\bbox \tau}_i\cdot {\bbox \tau}_j$,
${\bbox \sigma}_i\cdot{\bbox \sigma}_j$, ${\bbox \sigma}_i\cdot{\bbox
\sigma}_j {\bbox \tau}_i\cdot {\bbox \tau}_j$, $S_{ij}$, and
$S_{ij}{\bbox \tau}_i\cdot {\bbox \tau}_j$, and the $f^p_{ijk}$ is an
operator-independent three-body correlation.  The six radial functions
$f_{ss}(r)$ and $u_{p=2,6}(r)$ are obtained from two-body
Euler-Lagrange equations with variational parameters as discussed in
detail in Ref.~\cite{W91}.  They are taken to be the same in the
p-shell nuclei as in $^4$He, except that the $u_{p=2,6}(r)$ are forced
to go to zero at large distance by multiplying in a cutoff factor,
$\Big[1+{\rm exp}[-R_u/a_u]\Big]/\Big[1+{\rm exp}[(r-R_u)/a_u]\Big]$,
with $R_u$ and $a_u$ as variational parameters.  The $f_{sp}$
correlation is constructed to be similar to $f_{ss}$ for small
separations, but goes smoothly to a constant of order unity at large
distances ($r>5$ fm):
\begin{equation}
f_{sp}(r) = 
\Bigg[ a_{sp} + \frac{b_{sp}}{1+{\rm exp}[(r-R_{sp})/a_{sp}]} \Bigg] f_{ss}(r) 
+ c_{sp}(1-\exp[-(r/d_{sp})^2]) \ , \\
\end{equation}
where $a_{sp}$, $b_{sp}$, etc. are additional variational parameters.
The $f_{pp}(r)$ correlation in the mass-seven nuclei is the same as
the $f(r_{ij})$ correlation in the trinucleon, so that when the three
p-shell nucleons are all far from the s-shell core, they look very
much like a trinucleon.

These choices for $f_{ss}$, $f_{sp}$, $f_{pp}$, and $u_{p=2,6}$
guarantee that when the three p-shell particles are all far from the
s-shell core, the overall wave function factorizes as:
\begin{equation}
\Psi_T \rightarrow
[f_{sp}(r_{\alpha\tau})]^{12}[\phi^{LS[n]}_{p}(r_{\alpha \tau})]^3
\psi_{\alpha} \psi_\tau \ ,
\label{eqn:asymptote}
\end{equation}
where $\psi_{\alpha}$ is the variational $^4$He wave function,
$\psi_\tau$ is the variational trinucleon wave function, and ${\bf
r}_{\alpha\tau}$ denotes the separation between the centers of mass of
the $\alpha$ and trinucleon clusters.  Provided that
$[f_{sp}(r_{\alpha\tau})]^{12}$ goes to a constant quickly enough and
smoothly enough, the long-range correlation between clusters is
proportional to $[\phi^{LS[n]}_{p}(r_{\alpha \tau})]^3$.

The single-particle functions $\phi^{LS[n]}_{p}(R_{\alpha k})$ describe
correlations between the s-shell core and the p-shell nucleons, and
have been taken in previous work~\cite{ppcpw97} to be solutions of a
radial Schr\"odinger equation for a Woods-Saxon potential and unit
angular momentum, with energy and Woods-Saxon parameters determined
variationally.  It is important for low-energy radiative captures that
these functions reproduce faithfully the large-separation behavior of
the wave function, because the matrix elements receive large
contributions at cluster separations greater than 10 fm.  In fact, at
20 keV, more than 10\% of the cross section for \hag\ comes from
cluster separations beyond 20 fm.  At these distances, well outside
the nuclear interaction distance, the clusterization with the lowest
cluster-separation energy should be the most important.  We have
therefore modified the bound-state wave functions for the capture
calculation to enforce cluster-like behavior, matching laboratory
cluster separation energies, when the three p-shell nucleons are all
far from the s-shell core.

In general, for light p-shell nuclei with an asymptotic two-cluster
structure, such as $\alpha d$ in $^6$Li or $\alpha t$ in $^7$Li, we want the 
large separation behavior to be
\begin{equation}
\label{eqn:asymptotic}
[\phi^{LS[n]}_{p}(r\rightarrow\infty)]^n \propto W_{km}(2\gamma r)/r,
\end{equation}
where $W_{km}(2\gamma r)$ is the Whittaker function for bound-state wave 
functions in a Coulomb potential (see below) and $n$ is the number of 
p-shell nucleons.  We achieve this by solving the equation
\begin{equation}
\Bigg[ -\frac{\hbar^2}{2 \mu_{41}} \Bigg( \frac{d^2}{dr^2}
-\frac{\ell(\ell+1)}{r^2} \Bigg) +V(r)+\Lambda(r) \Bigg]
r\phi^{LS[n]}_{p}(r) = 0,
\end{equation}
with $\ell=1$, $\mu_{41}$ the reduced mass of one nucleon against
four, and $V(r)$ a parametrized Woods-Saxon potential plus Coulomb term:
\begin{equation}
V(r) = \frac{V_0}{1+{\rm exp}[(r-R_0)/a_0]} + \frac{2(Z-2)}{n} \frac{e^2}{r} F(r) \ .
\end{equation}
Here $V_0$, $R_0$, and $a_0$ are variational parameters, $(Z-2)/n$ is
the average charge of a p-shell nucleon, and $F(r)$ is a form factor
obtained by folding $\alpha$ and proton charge distributions together.
The $\Lambda(r)$ is a Lagrange multiplier that enforces the asymptotic
behavior at large $r$, but is cut off at small $r$ by means of a
variational parameter $c_0$:
\begin{equation}
\Lambda(r) = \lambda(r) \Big[ 1-{\rm exp}\Big(-(r/c_0)^2\Big) \Big] \ .
\end{equation}
The $\lambda(r)$ is given by
\begin{equation}
\lambda(r) = \frac{\hbar^2}{2 \mu_{41}} \Bigg[\frac{1}{u_L} \frac{d^2 u_L}{dr^2}
- \frac{2}{r^2} \Bigg] - \frac{2(Z-2)}{n} \frac{e^2}{r} \ ,
\end{equation}
where $u_L$ is directly related to the Whittaker function (solution to
the radial Schr\"odinger equation for negative-energy states in a
purely Coulomb potential):
\begin{equation}
u_L/r = (W_{km}(2\gamma r)/r)^{1/n} \ .
\end{equation}
Here $\gamma^2 = 2\mu_{4n} B_{4n}/\hbar^2$, with $\mu_{4n}$ and
$B_{4n}$ the appropriate two-cluster reduced mass and binding energy,
$k = -2(Z-2) e^2 \mu_{4n}/\hbar^2 \gamma$, and $m = L+\case{1}{2}$.

For the $^7$Li ground state, the largest contribution has $B_{43} =
2.47$ MeV (binding energy of $^7$Li relative to $\alpha$ and $t$
clusters) and $L =1$ corresponding to the asymptotic $P$-wave of the
$^7$Li ground state, or amplitude $\beta_{1\frac{1}{2}[3]}$ in
Eq. (\ref{eqn:jastrow}).  None of the other possible amplitudes
$\beta_{LS[n]}$ correspond to asymptotic $\alpha t$ clusterizations.
However, there is no reason for them not to be present in compact
configurations of the nucleons.  Including such components in the wave
functions improves the binding energies of the mass-seven bound states
by about $0.2$ MeV.  The asymptotic forms of $\phi^{LS[n]}_{p}(r)$ in
the lower-symmetry channels are set to match the threshold for $^7{\rm
Li} \rightarrow\ ^6{\rm Li} + n$.  Analogous descriptions hold for the
other bound states (the $^7$Li excited state and the two $^7$Be bound
states), with the appropriate thresholds substituted for $B_{43}$.

The $^7$Be ground and first excited state Jastrow functions have been
treated in previous development of the variational Monte Carlo wave
functions\cite{ppcpw97} as the isospin rotations ($T_3=+1/2$ instead
of $-1/2$) of the corresponding $^7$Li shell-model-like wave
functions.  In this work, the $^7$Li and $^7$Be bound state Jastrow
functions also differ by the choices of $B_{4n}$ for the asymptotic
cluster behavior of the wave functions, which match the cluster
breakup thresholds as described above in each case.  This choice of
$B_{4n}$ has the result that in configurations where the p-shell
nucleons are far from the s-shell core, the energy is the sum of the
Coulomb potential, the kinetic energy contributed by the
$\phi_p^{LS[n]}$, and the cluster binding energies (well-reproduced
because the core resembles an alpha particle and the p-shell has been
constructed to resemble a trinucleon).  Because the $\phi_p^{LS[n]}$
matches the laboratory binding energy for the known Coulomb potential,
the local energies at large particle separations match the known
binding energies in the mass-seven nuclei.  This agreement has been
confirmed numerically.

In Refs.~\cite{ppcpw97,APW95} the $f^{sss}_{ijk}$ three-body
correlation of Eq.(\ref{eqn:jastrow}) was a valuable and
computationally inexpensive improvement to the trial function, but no
$f^{ssp}_{ijk}$ or $f^{spp}_{ijk}$ correlations could be found that
were of any benefit.  However, for the types of wave functions used
here, it is found that the correlation
\begin{equation}
f^{spp}_{nij} = 1 + q_1 [f_{ss}(r_{ij})/f_{pp}(r_{ij}) - 1] 
                        {\rm exp}[-q_2(r_{ni}+r_{nj})] \ ,
\end{equation}
with $i,j$ labels of p-shell nucleons and $q_{1,2}$ as variational
parameters, is very useful \cite{nws}.  It effectively alters the
central pair correlations between pairs of p-shell nucleons from their
trinucleon-like forms to be more like the pair correlations within the
s-shell when the two particles are close to the core.  This
correlation improves the binding energy by $\approx 0.25$ MeV in
$^7$Li.

The authors of Refs.~\cite{ppcpw97,WPCP00} reported energies both for
the trial function $\Psi_T$ of Eq.(\ref{eqn:trial}), and for more
sophisticated variational wave functions, $\Psi_V$, which add two-body
spin-orbit and three-body spin- and isospin-dependent correlation
operators.  The $\Psi_V$ gives improved binding compared to $\Psi_T$
in both the mass-four and the mass-seven wave functions considered,
but is significantly more expensive to construct because of the
numerical derivatives required for the spin-orbit correlations.  In
the case of an energy calculation, the derivatives are also needed for
the evaluation of $L$-dependent terms in AV18, so the cost is only a
factor of two in computation.  However, for the evaluation of other
expectation values the relative cost increase is $\approx 6A$.  Thus
in the present work we choose to use $\Psi_T$ for non-energy
evaluations; this proved quite adequate in studies of $^6$Li form
factors~\cite{WS98} and of the six-body radiative capture \dalphg
\cite{nws}.

The variational Monte Carlo (VMC) energies and point proton RMS radii
obtained with $\Psi_T$ are shown in Table~\ref{tab:energy} along with
the results of essentially exact Green's function Monte Carlo (GFMC)
calculations \cite{ppcpw97,WPCP00,pieper00} and the experimental
values.  We note that the underbinding of the $A=7$ nuclei in the GFMC
calculation arises from the AV18/UIX model and not the many-body
method; it can be improved by the introduction of more sophisticated
three-nucleon potentials~\cite{PPWC01}.

\begin{table}
\caption{Calculated VMC, GFMC, and experimentally measured energies,
point proton RMS radii, and quadrupole moments of $^3$H, $^3$He $^4$He
and bound states of $^7$Li and $^7$Be. Numbers in parentheses are
Monte Carlo statistical errors.}
\begin{tabular}{ccddd}
Nucleus & Observable & {VMC $\Psi_T$} & GFMC & Experiment \\
\hline
$^3$H  & E           &  --8.15(1)     & --8.47(1) &  --8.48  \\
       & $\langle r^2_p \rangle^{1/2}$ 
           &      1.60(1)             &   1.59(1) &    1.60 \\
\\
$^3$He  & E          &  --7.39(1)     &        &  --7.72 \\
       & $\langle r^2_p \rangle^{1/2}$
           &    1.73(1)    &              1.73(1) &      1.77 \\
\\
$^4$He & E           & --26.89(3)     & --28.34(4) & --28.30 \\
       & $\langle r^2_p \rangle^{1/2}$
           &    1.48(1)    &    1.45(1)        &    1.48 \\
\\
$^7$Li & E           & --31.26(8)    & --37.78(14)& --39.24 \\
       & $\langle r^2_p \rangle^{1/2}$
           &    2.30(1)  &              2.33(1)&    2.27  \\
       & Q &   --3.7(2)  &             --4.5(2)&  --4.06 \\
\\
$^7$Be & E           & --29.55(8)    & --36.23(14)& --37.60  \\
       & $\langle r^2_p \rangle^{1/2}$
           &    2.41(1)  &              2.52(1)&           \\
       & Q &   --5.9(3)  &             --7.5(3)&           \\
\\
$^7{\rm Li}^*$  & E  & --31.37(8)     & --37.53(15)& --38.77 \\
                & $\langle r^2_p \rangle^{1/2}$
           &    2.35(2)  &    2.35(2)          &            \\
\\
$^7{\rm Be}^*$ & E & --29.70(8)      & --36.01(15)& --37.17 \\
                & $\langle r^2_p \rangle^{1/2}$
                &    2.46(2)  &        2.54(2) &            \\
\end{tabular}
\label{tab:energy}
\end{table}

Although the present variational trial functions with the imposed
Coulomb asymptotic correlations produced a variational improvement in
the case of $^6$Li, they give approximately the same energies as the
older shell-model-like correlations of Refs.~\cite{ppcpw97,WPCP00} for
$^7$Li and $^7$Be.  Unfortunately, because the variational energies of
the $A=7$ bound states are not below those of separated alpha and
trinucleon clusters, it is possible to lower the energy significantly
by making the wave functions more diffuse.  Therefore, the variational
parameters were constrained to give RMS charge radii which agree
reasonably with experiment, as seen in Table~\ref{tab:energy}.  Given
$\phi_p^{LS[n]}$, the easiest way to constrain the RMS radius in the
variational procedure is to choose the parameters of the $f_{sp}$
correlations between s- and p-shell nucleons to adjust the probability
of finding the p-shell nucleons far from the s-shell core.  There was
considerable freedom in the specific form of these correlations as
long as asymptotic properties of the wave functions were not being
tested, because of the insensitivity of energy expectation values to
the tails of the wave functions.  However, the form of the correlation
in the $\alpha\tau$ channel depends on $f_{sp}^{12}$, as seen in
Eq. (\ref{eqn:asymptote}), so the large-cluster-separation parts of
the wave function are very sensitive to the choice of $f_{sp}$.  (The
twelfth power arises because there are four particles in the s-shell
and three in the p-shell, and thus $4\times 3=12$ sp pairs.)  Prior to
this work, VMC wave functions had cluster distributions that dropped
by an extra factor of two beyond 5 fm, relative to the drop expected
on the basis of the clusterization arguments given above.  The extra
drop had two closely-related effects: the asymptotic normalization
coefficients (see below) for two-cluster breakup were too small, and
so were the cross sections which we initially computed from them -- by
a factor of about two.  The present wave functions perform more poorly
than previous VMC wave functions on the ordering of the mass-seven
bound states (a perennial difficulty for both VMC and GFMC because of
the close spacing of the states), but they give larger asymptotic
normalizations at a reasonable cost in binding energy.  The
relationship between nuclear size (as measured by quadrupole moments)
and cross sections for direct radiative captures has been noticed
before, and applied usefully both to these \cite{kajino86} and other
reactions, most notably $^7{\rm Be}(p,\gamma)^8{\rm B}$
\cite{csoto98}.  We note that our $S(0)$ and quadrupole moments for
both mass-seven systems fit the general trends shown in
Ref. \cite{csoto00}.

\begin{figure}
\centerline{\epsfig{file=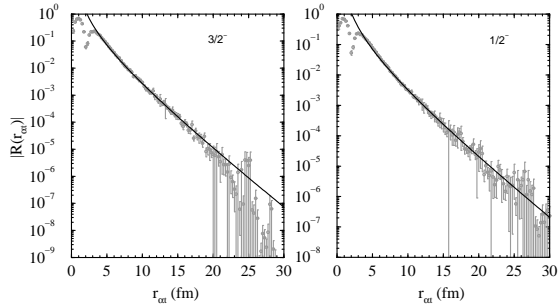,height=7.2cm,angle=270}}
\caption{Monte Carlo samples of the radial two-cluster $\alpha t$
distribution functions in $^7$Li ground state ($J^\pi=3/2^-$, left
panel) and first excited state ($J^\pi=1/2^-$, right panel), with
error estimates.  The solid curve is the expected Whittaker-function
asymptotic form of the overlap, normalized as in Table
\protect\ref{tab:anc} to match the $\alpha\tau$ distributions at 7--10
fm.}
\label{fig:lirls}
\end{figure}

\begin{figure}
\centerline{\epsfig{file=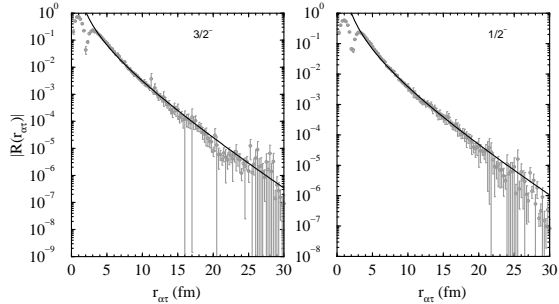,height=7.2cm,angle=270}}
\caption{Same as Fig. \protect\ref{fig:lirls}, but for radial
two-cluster $\alpha ^3{\rm He}$ distribution functions in $^7$Be bound
states.}
\label{fig:berls}
\end{figure}

The asymptotic two-cluster behavior of the seven-body wave functions
can be studied by computing the two-cluster $\alpha \tau$ distribution
functions, $\langle{\cal A}\psi_{\alpha}\psi_t^{m_t},{\bf r}_{\alpha
t}\mid\psi_{\rm Li}^{m_7}\rangle$ for $^7$Li, and its analogs for the
other seven-body bound states we consider.  These functions are
described in Ref.~\cite{FPPWSA96}.  They can be expressed in terms of
Clebsch-Gordan factors, spherical harmonics, and the radial functions
$R(r_{\alpha \tau})$ plotted in Figs.~\ref{fig:lirls} and
\ref{fig:berls}.  At large $r_{\alpha \tau}$, $r_{\alpha
\tau}R(r_{\alpha \tau})$ should be proportional to a Whittaker
function, as described above.  The proportionality constant is the
asymptotic normalization constant $C_1$.  We have extracted $C_1$ from
the overlap functions by a least-squares procedure, matching them to
the appropriate Whittaker functions.  We find that for all the
mass-seven wave functions considered here, $C_1$ becomes asymptotic at
$r_{\alpha \tau} \approx $ 7--9 fm.  In Table \ref{tab:anc} we present
asymptotic normalization coefficients for the $A=7$ bound states,
based on fitting overlaps in the region 7--10 fm as shown in
Figs. \ref{fig:lirls} and \ref{fig:berls}.

\begin{table}
\caption{Asymptotic normalization coefficients (in ${\rm fm}^{-1/2}$)
for the overlap between mass-seven bound states and $\alpha\tau$
clusterization, computed from the VMC wave functions.  Best available
estimates are presented for comparison on the lower line.}
\label{tab:anc}
\begin{tabular}{cccccc}
& $^7$Li & $^7{\rm Li}^*$ & $^7$Be & $^7{\rm Be}^{*}$ \\
\hline
VMC & $3.4\pm 0.1$ & $2.65\pm 0.10$ & $3.55\pm 0.15$ & $2.9\pm 0.1$\\
literature & $3.55\pm 0.27$ & 3.14 & 4.79 & 4.03\\
\end{tabular}
\end{table}

Our values of $C_1$ for the mass-seven nuclei tend to be somewhat
smaller than the values found in the
literature~\cite{brune99,igamov97}.  Only in the case of the $^7$Li
ground state are experimental determinations of the $C_1$ of
reasonable quality.  Brune et al. \cite{brune99} find a ``world
average'' of $3.55\pm 0.27\ {\rm fm}^{-1/2}$, relying mainly on
theoretical models\cite{mohr93,kajino86}, and giving less weight to
the partially experimentally-based evaluations of
Igamov\cite{igamov97}.  For the other states, we rely on the Igamov et
al. \cite{igamov97} extraction of ANCs from Kajino's RGM calculations
with the MHN potential\cite{kajino86}.  These calculations match the
radiative capture data for \tag\ very well in both energy dependence
and normalization.  However, Igamov et al. report ``nuclear vertex
constants,'' which differ from ANCs by prefactors whose definitions
are ambiguous in the literature; the numbers presented in the second
row of Table \ref{tab:anc} should be used with caution.  The present
results for the ANCs are also subject to correlated uncertainties
characteristic of the Monte Carlo integration method used to compute
the overlaps, and the uncertainties are therefore difficult to
estimate reliably (as discussed in more detail with regard to matrix
element densities in Sec. \ref{sec:integration} below).

As noted in the Introduction, low-energy direct captures can be
treated to good approximation by considering only cluster separations
beyond a few fm, and using only the longer-range parts of the bound-
and initial-state wave functions to compute matrix elements.  This
approach requires the provision of ``spectroscopic factors,'' or more
precisely, the ANCs discussed above.  Because we have imposed the
condition that the large-cluster-separation part of the ground state
match its expected form, it is true that for our cross section
calculations, the VMC method provides $C_1$.  However, it also
provides the inner few fm of the wave function, which requires some
model of what is going on inside the nuclear interaction radius, and
may be important for understanding differences in the logarithmic
derivatives of the $S$-factor found in various theoretical studies.

\subsection{Scattering states}

The initial-state wave functions are taken to be elastic-scattering
states of the form
\begin{equation}
\label{eqn:scatstate}
|\psi_{\alpha\tau}; LSJM \rangle = 
{\cal A}\left\{\phi_{\alpha\tau}^{JL}(r_{\alpha\tau})Y_{LM_L}({\bf\hat{r}}_{\alpha\tau})
\prod_{ij}G_{ij}|\psi_\alpha\psi_\tau^{m_S}\rangle\right\}_{LSJM},
\end{equation}
where curly braces indicate angular momentum coupling,  $\cal{A}$
antisymmetrizes between clusters, $\psi_\alpha$ is the $^4$He ground
state, and $\psi_\tau^{m_S}$ is the trinucleon ground state in spin
orientation $m_S$.

The $G_{ij}$ are identity operators if the nucleons $i$ and $j$ are in
the same cluster.  Otherwise, they are a set of both central and
non-central pair correlation operators which introduce distortions in
each cluster, under the influence of individual nucleons from the
other cluster.  They are derived from solutions for nucleon-nucleon
correlations in nuclear matter \cite{lagaris81}, and become the
identity operator at pair separations beyond about 2 fm.  (These
correlations have been included in the definition of the overlap
functions shown in Figs. \ref{fig:lirls} and \ref{fig:berls}.)

The correlations $\phi_{\alpha\tau}^{JL}$ are derived
phenomenologically.  The variational seven-body bound-state wave
functions do not give the correct energies with respect to cluster
breakup, so we do not expect to be able to use the variational
technique to solve for these correlations.  Instead, we generate the
$\phi_{\alpha\tau}^{JL}$ as solutions to Schr\"odinger equations, from
cluster-cluster potentials that describe phase shifts of $\alpha$-\h3
and $\alpha$-\he3 scattering as scattering of point particles.
Because of the small amount of available laboratory data and the large
amount of work that has already been put into generating potentials
that reproduce them, we take cluster-cluster potentials from the
literature \cite{dubovichenko95,kim81,buck85,buck88,langanke86}.  We
now point out the main features of these potentials.

We treat all of these models as (and many have been explicitly
constructed as) descriptions of both the $\alpha t$ and
$\alpha\,^3{\rm He}$ systems, with appropriate Coulomb potentials and
laboratory masses.  Each of the potentials we use to generate the
$\phi_{\alpha\tau}^{JL}$ has a deep, attractive central term, and a
spin-orbit term.  The spin-orbit terms are constrained mainly by the
spacing between the $P_{3/2}$ and $P_{1/2}$ bound states, and between
the resonances at 2.16 and 4.21 MeV in $F_{7/2}$ and $F_{5/2}$ $\alpha
t$ scattering, respectively (as well as their analogs in $\alpha\,
^3{\rm He}$ scattering).  The other interesting feature in the
scattering of the odd partial waves is the apparent hard-sphere
behavior of the phase shifts in $P$-wave scattering.  This comes about
because the wave functions between the clusters must respect the Pauli
principle by allowing antisymmetrization of nucleon wave functions
between clusters.  In the $P$ waves, this takes the form of wave
functions that have a single node whose location is almost independent
of scattering energy, and it therefore gives rise to phase shifts
which look like scattering from a hard sphere \cite{zaikin71}.  (See
Figs. \ref{fig:lirls} and \ref{fig:berls}, where the steep dips in
absolute values of the cluster distributions at 2 fm correspond to
nodes).  In the cluster-cluster potentials, this requires that the
central term be large and negative, so that the ground state wave
function has one node.  Of course, a more tightly-bound ``forbidden''
state with zero nodes exists for such a potential, but it does not
allow antisymmetrization of the nucleon wave functions.  We note that
the requirement of a particular nodal structure only models
approximately the effects of the Pauli principle on the inter-cluster
correlations.  We did not use potentials that enforced the hard-sphere
behavior with repulsive short-range terms \cite{shulgina96}.

\begin{figure}
\centerline{\epsfig{file=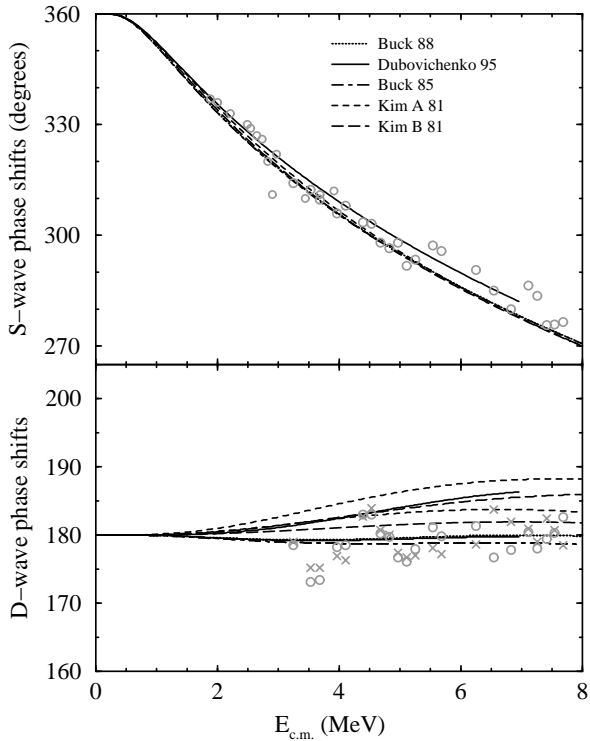,height=10cm,angle=0}}
\caption{Phase shifts produced by the potentials used to generate the
$\phi_{\alpha\tau}^{JL}$.  Data are taken from
Refs. \protect\cite{boykin72,hardy72}; in the lower panel, $\times$
denotes $J^\pi=\frac{5}{2}^+$ phase shifts, $\circ$ denotes
$J^\pi=\frac{3}{2}^+$ phase shifts.  Potentials are taken from
Refs. \protect\cite{dubovichenko95,kim81,buck85,buck88}.}
\label{fig:phaseshifts}
\end{figure}

Because of the requirements of the Pauli principle, one also expects
different potentials to describe the odd- and even-parity scattering.
The even-parity phase shifts are unfortunately lacking in details
which models must match (see Fig. \ref{fig:phaseshifts}), beyond the
apparent hard-sphere scattering in the $S$-wave data, corresponding to
the Pauli-required minimum of two nodes in the wave function
\cite{zaikin71}.  The measured $D$-wave phase shifts are even worse,
being consistent with zero (or $180^\circ$, with the knowledge that
there must be at least one node in the wave function) throughout the
region below 8 MeV for both $\alpha$-t and $\alpha$-$^3{\rm He}$
systems \cite{spiger67,ivanovich68,boykin72,hardy72}.  This lack of
features is particularly unfortunate in that by far the most important
reaction mechanisms for radiative capture in this system are $E1$
captures from $S$- and $D$-wave scattering states.  The most stringent
tests of any of the $D$-wave potentials have been comparison of their
phase shifts with the results of more elaborate (RGM) theoretical
models.

Finally, we note the difficulty of reproducing the published models,
which is due to omitted descriptions of details of the potentials,
particularly handling of the Coulomb potential at short range.  Rather
than guess how to fix up each potential, we restrict attention to
potentials that allowed good descriptions of the low-energy scattering
on the first try (with no short-range cutoff in the $1/r$ Coulomb term
unless explicitly described by the potential's authors).  Relatively
small differences in the cluster-cluster potentials are directly
connected to the size of the scattering wave function at energies less
than 1 MeV and cluster separations less than 20 fm, resulting directly
in differences in the normalization of our computed capture cross
sections from one potential to the next.  However, it is possible to
eliminate the worst potentials on the basis of the low-energy phase
shifts; potentials which underpredict the phase shifts also produce
radiative capture cross sections that are too low by as much as a
factor of two, relative to those generated from other cluster-cluster
potentials.  Potentials which were created for use in orthogonal
cluster models (as opposed to simple cluster models) were eliminated
from application to our problem on this criterion.

\section{Operators}
\label{sec:operators}

The cross sections were computed by a multipole expansion of the
electromagnetic current operator \cite{enchilada},

\begin{equation}
\sigma(E_{\rm c.m.})= \sum_{LSJ_iJ_f\ell} \frac{8\pi}{2J_f+1} 
\frac{\alpha}{v_{\rm rel}}
\frac{q}{1+q/m_{\rm 7}} \Bigg[ \left|E^{LSJ_iJ_f}_\ell(q)\right|^2
+\left|M^{LSJ_iJ_f}_\ell(q)\right|^2 \Bigg] \ ,
\label{eqn:multipole}
\end{equation}
where $\alpha$ is the fine structure constant ($\alpha=e^2/\hbar c$),
$v_{\rm rel}$ is the $\alpha$$\tau$ relative velocity, $m_7$ is the
mass of the final state, and $E^{LSJ_iJ_f}_\ell(q)$ and
$M^{LSJ_iJ_f}_\ell(q)$ are the reduced matrix elements (RMEs) of the
electric and magnetic multipole operators with multipolarity $\ell$
connecting the scattering states in channel $LSJ_i$ to bound states of
$^7$Li or $^7$Be with angular momentum $J_f$.  The center-of-mass
energy of the emitted photon is given by

\begin{eqnarray}
q&=&m_{\rm 7}\Bigg[ -1 + \sqrt{1+\frac{2}{m_7}
(m_\tau+m_\alpha-m_7+E_{\rm c.m.})} \Bigg], \nonumber\\
&\simeq& m_\alpha + m_\tau - m_7 + E_{\rm c.m.}
\end{eqnarray}
where $m_\tau$, $m_\alpha$, and $m_7$ are the rest masses of the
trinucleon, $^4$He, and the appropriate $^7$Li or $^7$Be state,
respectively.  The astrophysical $S$-factor is then related to the
cross section via
\begin{equation}
S(E_{\rm c.m.}) = E_{\rm c.m.}\, \sigma(E_{\rm c.m.})\, {\rm exp}( 2 Z_1 Z_2\, \pi
\, \alpha/\hbar v_{\rm rel}) \ ,
\end{equation}
where $Z_1$ and $Z_2$ are the charges of the two initial-state nuclei.

The dominant reaction mechanism, for creation of both the excited
state and the ground state, is an $E1$ (electric dipole) transition
from an $S$-wave scattering state to a bound state.  Capture into the
excited state is followed immediately by electromagnetic decay to the
ground state, so the total cross section of interest for astrophysics
is the sum of the cross sections for captures into the two states.  At
energies above about 500 keV, $E1$ capture from $D$ waves becomes
important.  We computed transitions originating from scattering states
with orbital angular momentum $L=0,1,2$ and 3, via $M1,M2,E1,E2,$ and
$E3$ transitions, and found that up to the 0.1\% level, only $E1$
captures originating from $S$- and $D$-wave scattering states matter
at energies below 1.5 MeV.

With the exception of the $E1$ term, all RMEs were computed in the
standard long-wavelength approximation (LWA), keeping only the
lowest-order term in photon wavenumber of the modified spherical
Bessel functions appearing in the RME integrals.  The LWA is valid to
reasonable accuracy because at the low energies under consideration
($E<1$ MeV), the ratio of system size to photon wavelength is less
than $10\ {\rm fm} / 200\ {\rm fm} = 0.05$.

In a previous study \cite{nws}, we developed code to examine the
isospin-forbidden $E1$ transition in \dalphg.  We have applied this
code to compute corrections to the LWA for the $E1$ transitions under
consideration here.  An examination of these corrections for the
mass-seven system is in principle of interest for the problem of
extrapolating cross sections to low energies.  However, we find that
all but one of them provide contributions of less than 0.05\% of the
total cross section.  (See Ref. \cite{nws} for a list and detailed
discussion of the corrections we applied, which extend to third order
in $q$.)  We do not actually compute the largest correction, which is
the ``center-of-energy'' correction.  This correction arises because
potentials and kinetic energies should be included in the definition
of the center-of-momentum frame, but have not been\cite{boosts}.  The
center-of-energy correction becomes important when the leading order
LWA operator vanishes, as in \dalphg, but it should amount to only
+2.4\% of the $E1$ cross section in \tag\ and +3.1\% of the $E1$ cross
section in \hag.  We did not compute center-of-energy corrections (or
include an estimate of them in the results presented below) because of
the extra computation necessary to find energies during the capture
calculation.  Their omission is not serious because 1) their effect is
to change the normalization, not the energy dependence of the $E1$
cross section, 2) our model calculation is only accurate to about
5--10\% at best, and 3) the above estimates of the size of the effect
should be quite accurate, being based only on the differences between
using nuclear masses and using integer multiples of the mean nucleon
mass in the factor $[(Z_2 m_1-Z_1 m_2)/(m_1+m_2)]^2$ in the LWA cross
section.

\subsection{Matrix element integration}
\label{sec:integration}

Actual computation of the matrix elements was performed with a
modified version of the code described in Ref.~\cite{nws}, which is
itself a modified version of a code developed to compute energies and
other properties of light nuclei~\cite{ppcpw97} for variational
calculations.  The method used to integrate over nucleon
configurations is the Metropolis Monte Carlo algorithm, with a weight
function proportional to the bound state wave function involved in the
computed transition.  As discussed below, this weight function was
chosen to reflect in general detail the form of the matrix element
integrands, and to obtain significant numbers of Monte Carlo samples
over a broad range of cluster separations.  The final calculation
consisted of $10^6$ samples for each transition.

We have applied the approach of splitting the calculation into
energy-dependent and energy-independent parts, as in Ref.~\cite{aps},
so that the reduced matrix elements of Eq. (\ref{eqn:multipole}) are
written
\begin{eqnarray}
\label{eqn:integrand}
T^{LSJ_iJ_f}_\ell(q)&&=\frac{\sqrt{2J_f+1}}{<J_iM_i,\ell\lambda|J_fm_f>}\\
&&\times\int_0^\infty
dx\, x^2 \, \phi_{\alpha\tau}^{J_iL}(x)\langle \psi_{7}^{J_fm_f}|
T_{\ell \lambda}(q) {\cal A}\left\{\delta(x-r_{\alpha\tau} )
Y_L^{M_L}({\bf\hat{r}}_{\alpha\tau}))
\prod_{ij}G_{ij}|\psi_\alpha\psi_\tau^{m_S}\rangle\right\}_{LSJ_iM_i}
\end{eqnarray}
and computed using photon polarization $\lambda=+1$ for the multipole
operators $T_{\ell \lambda}(q)$.  The delta function is applied by
accumulating the Monte Carlo integral in radial bins of thickness
$0.25$ fm.  The final integration over $x$ is performed by inserting
the appropriate dependences on photon energy in each term (since the
LWA expansion is in powers of energy, this dependence may be taken out
of the integral), and computing $\phi_{\alpha\tau}^{JL}$, at each
energy.  This allows the time-consuming Monte Carlo integration to be
performed only once for each partial wave and operator, so that
computation of RMEs for many energies is relatively inexpensive in
computer time.  After initially setting up the code and checking that
selection rules were satisfied, we did not explicitly compute RMEs for
parity-forbidden operators.

\begin{figure}
\centerline{\epsfig{file=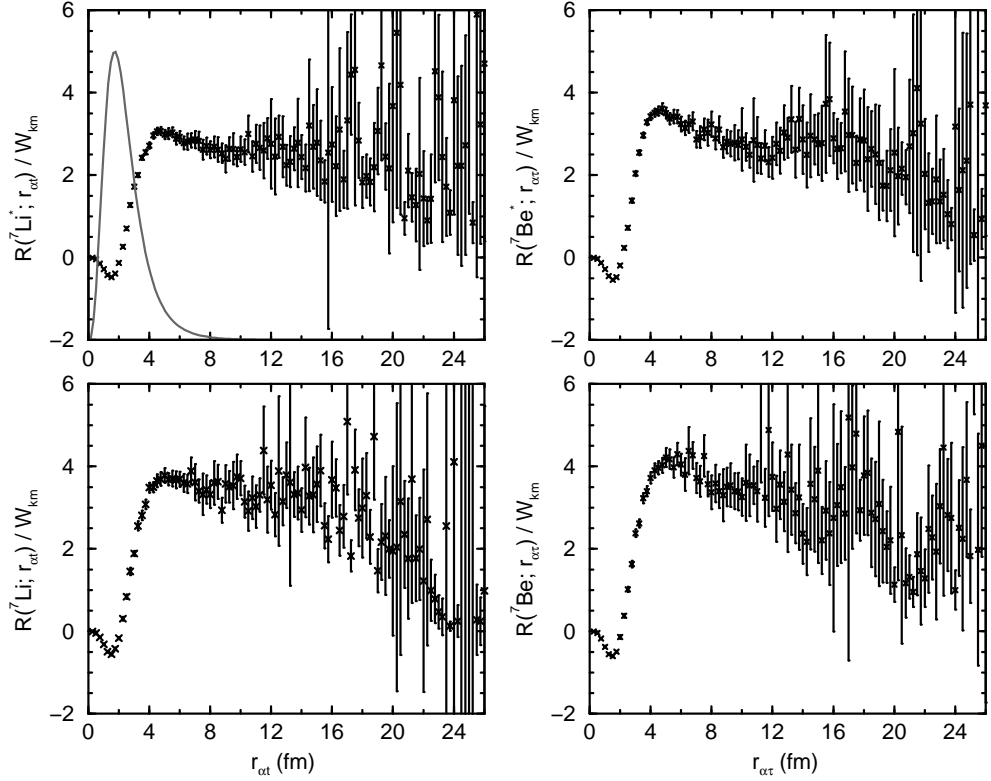,height=13cm,angle=270}}
\caption{Ratios of the computed radial overlap functions of
Figs. \protect\ref{fig:lirls} and \protect\ref{fig:berls} to their
imposed asymptotic forms, based on $10^6$ particle configurations.
These ratios should be equal to the asymptotic normalization constant
for the appropriate wave function beyond about 7 fm.  Superimposed on
the ratios for the $^7$Li excited state is the corresponding
distribution of Monte Carlo samples, essentially identical for all
four overlap functions.}
\label{fig:anc}
\end{figure}

At cluster separations beyond about 10 fm, the RME densities were
subject to considerable noise in the Monte Carlo sampling.  This is
because while Monte Carlo weighting schemes based on the ground state
give good sampling along directions other than the cluster separation
in the $3A$-dimensional configuration space, they provide small
numbers of samples at large $r_{\alpha\tau}$ (due to exponential decay
of the wave function at large distances).  VMC work usually uses
weighting based on the inner product in spin-isospin space of a
simplified bound state with itself.  Such weighting is good for
minimizing Monte Carlo variance of integrands that resemble the bound
state probability density closely, such as energies, but it does not
provide enough samples in the asymptotic region of the wave function
to do low-energy direct capture calculations.  Our previous paper
\cite{nws} used the square root of the usual weight function,
extending the Monte Carlo sampling out to large cluster separations,
but at the expense of greater sampling noise for a fixed number of
samples at fixed cluster separation.  In the mass-six problem, it was
straightforward to run the Monte Carlo integration of the RMEs until
the densities had ``converged'' to the expected asymptotic forms at
large cluster separation.  This required about $2\times 10^6$ total
configurations for a given RME.  At $A=7$, the spin-isospin space is
larger, so the code is slower by a factor of about 10, and it was only
practical to obtain $10^6$ samples for each RME with available
computing resources.  However, since the configuration space gains
three dimensions with the addition of a particle, $10^6$ samples do
not provide as thorough of a sample of the configuration space at mass
7 as at mass 6.  The result is at best a few-percent measure of the
asymptotic normalization at 7--10 fm.  Many samples are obtained at
larger cluster separation with the new weighting scheme, but the
samples beyond 15 fm appear to have correlated noise.  This is
presumably because the samples in question correspond to only a few
excursions of the sampling Markov chain into the region of large
cluster separation, and are therefore not very independent from each
other.  They tend to be either mostly high or mostly low relative to
the asymptotic forms explicitly built into the wave functions.  See
Fig. \ref{fig:anc} for illustration of these difficulties.  A much
larger sample (by a factor of 10) would be expected to exhibit much
less of this sort of correlated noise, but is prohibited by the large
amount of computer time that would require.

The matrix elements were therefore integrated out to 7 fm using the
Monte Carlo results for the integrand of Eq. (\ref{eqn:integrand});
integration beyond 7 fm was carried out using the known asymptotic
forms of the matrix element densities, normalized to the Monte Carlo
output by least-squares fitting at 7--10 fm.  This range was arrived
at by comparing results for the ANC of the $^7$Li ground state arising
from two different weighting schemes and varying numbers of samples.
Reasonably consistent agreement was found by fitting at 7--10 fm in
all cases.  The accuracy in the cross section then depends on how
accurately the ANCs can be determined in this region.

\section{Cross sections}
\label{sec:results}

\subsection{\tag}

\begin{figure}
\centerline{\epsfig{file=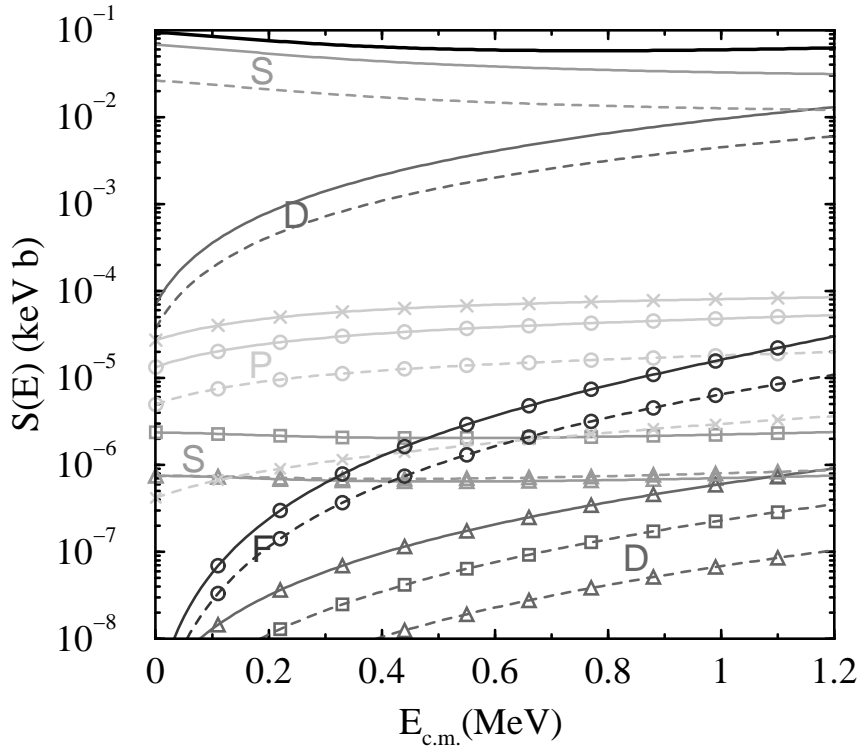,height=10cm,angle=0}}
\caption{Low-energy $S$-factor for \tag\ (thick solid line), and its
breakdown into contributions from various partial strengths, computed
using potential A of Kim et al. \protect\cite{kim81}.  Transitions to the
ground state are shown as solid lines, and transitions to the excited
state as dashed lines.  Labels indicate the initial state, and
additional symbols indicate multipole operator: no symbol, $E1$;
$\circ$, $E2$; $\times$, $M1$; $\Box $, $M2$; and $\triangle$,
order $q^3$ spin correction to $E1$ operator (LWA2 of
Ref. \protect\cite{nws}).}
\label{fig:tag}
\end{figure}

\begin{figure}
\centerline{\epsfig{file=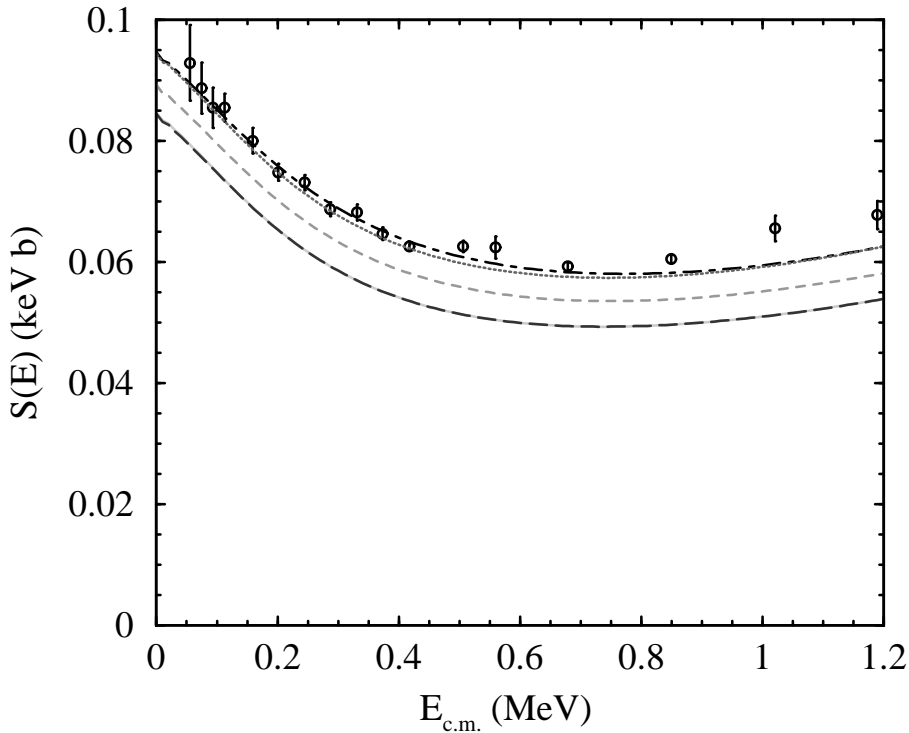,height=10cm,angle=0}}
\caption{Same as Fig. \protect\ref{fig:tag}, but showing only total
$S$-factors for several different potentials used to generate cluster
correlation functions $\phi_{\alpha t}^{JL}$: dot-dashed and dotted,
potentials A and B respectively of Kim et al.\protect\cite{kim81};
short-dashed, from Dubovichenko et al.\protect\cite{dubovichenko95},
solid, from Buck et al.  \protect\cite{buck85}; and long-dashed, from
Buck and Merchant \protect\cite{buck88}.  Data are from Brune et
al. \protect\cite{tag-brune}, and share a common 6\% normalization
uncertainty, not shown.}
\label{fig:tag-variation}
\end{figure}

The computed $S$-factor for \tag\ is shown broken down into
contributions from various terms of Eq. (\ref{eqn:multipole}) in
Fig. \ref{fig:tag}, and in comparison with laboratory data in
Fig. \ref{fig:tag-variation}.  The dominant processes are obviously
$E1$ captures, with large contributions from captures into both the
ground and excited states.  Contributions from other partial waves and
multipole operators are not present above the 1\% level.
Contributions from higher-order LWA corrections to the $E1$ operator
are less than 0.02\%.  Our calculations are therefore limited in
accuracy only by 1) the accuracy of the (bound- and scattering-state)
wave functions, and 2) Monte Carlo statistics.

The present calculation is the result of sampling $10^6$ points in the
seven-particle configuration space.  Formal error estimation on the
resulting cross sections is difficult because all 35 partitions of the
seven nucleons into alpha and triton clusters were computed for each
configuration.  This saves computation time, and enforces antisymmetry
of the initial state exactly; it also introduces correlations between
values of the operator densities (integrands of
Eq. (\ref{eqn:integrand})) at different cluster separations.  The
uncertainties in different transitions and partial waves are also
correlated, because they are based on the same random walk of particle
configurations.  There is also an uncertainty from the wave function
normalizations, because they are also derived from Monte Carlo
integrations.  This last uncertainty amounts to about 3\% in the cross
section.

We take the best indication of the Monte Carlo uncertainty to be the
formal uncertainties on the asymptotic normalizations of the Monte
Carlo matrix element densities, which we used to compute matrix
element contributions beyond cluster separations of 7 fm.  The
asymptotic forms of the matrix element densities are the
Whittaker-function asymptotic forms discussed above, multiplied by the
radial dependences of the electromagnetic multipole operators.  The
normalizations were fitted to the matrix element densities at cluster
separations of 7--10 fm.  Using fitted asymptotic forms amounts to
treating a large part of the matrix element as an external direct
capture, and it fixes two problems.  First, it removes the need for
large numbers of Monte Carlo samples in the remote tails of the wave
function.  Second, the asymptotic normalizations are found by a
weighted least-squares procedure, and formal error estimates on these
normalizations are possible.  In practice, the correlations between
matrix element densities produce reduced $\chi^2$ significantly less
than unity.  A common approach when confronted with such a problem in
experimental data is to assume that uncertainties have been
overestimated, and to reduce formal uncertainties accordingly.  We
have not done this, and we arrive at uncertainties of approximately
10\% in the cross section, based on the formal error estimates
(corresponding closely to the sizes of ANC errors in Table
\ref{tab:anc}).  More detailed analysis is problematic, and it is not
called for because the 10\% estimate is already larger than any other
contribution to the error budget.

The results themselves are best characterized in three ways:

1) Normalization.  Our results are lower than the data of Brune et
al. \cite{tag-brune} by 0--20\%, depending on the cluster potential.
Although other data exist
\cite{tag-holmgren,tag-schroeder,tag-griffiths,tag-burzynski,tag-utsunomiya},
those of Brune et al. are much more precise, and permit the best test
of our results.  Those data share a common 6\% normalization
uncertainty, not shown in Fig \ref{fig:tag-variation}.  The systematic
discrepancy between some of our results and the data is probably small
enough to ascribe to the combined uncertainties of the Monte Carlo
integration and of the data.  Taking the branching ratio to have its
experimental value of $R=0.453$, the computed $S$-factors for
transitions to the ground state match the measured low-energy
$S$-factors, while those for the excited state do not.  The
normalization is affected significantly by the choice of potential
used to generate the inter-cluster correlations $\phi_{\alpha t}^{JL}$
for the scattering states.  By applying five different potentials from
the literature as described above, we obtain a variation of $\pm 5\%$
in the $S(0)$ (total range for the five potentials) about a mean of
0.90 keV$\cdot$b -- a full range equal in size to the Monte Carlo
uncertainty.  Summarizing our results in one number, we obtain using
the inter-cluster potential A of Kim et al.\cite{kim81} (the best fit
to the $S$-wave phase shifts), $S(0)=0.095$ keV b, similar to other
estimates found in the literature.

\begin{figure}
\centerline{\epsfig{file=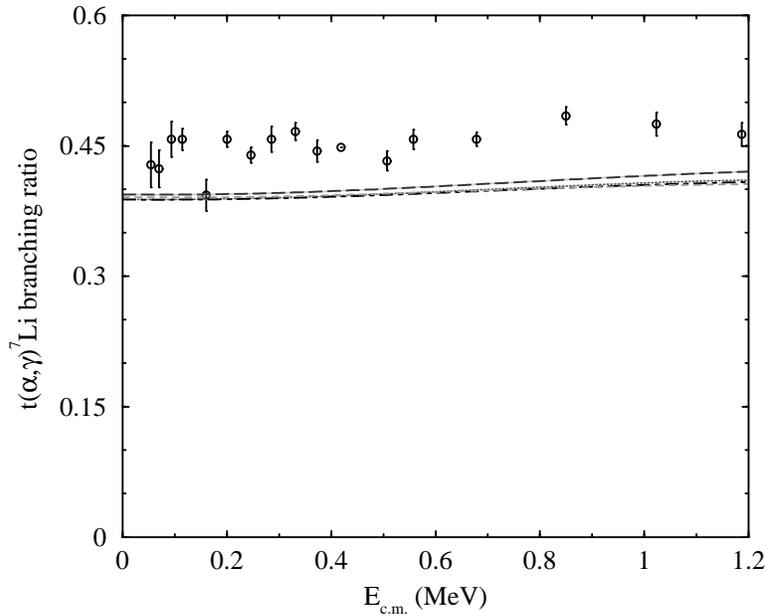,height=10cm,angle=270}}
\caption{Computed ratio of the cross section for capture into the
excited state to that for capture into the ground state, for \tag.
Symbols are the same as in Fig. \protect\ref{fig:tag-variation}, and
the data share a normalization uncertainty of 4\% (not shown).}
\label{fig:tag-branch}
\end{figure}

2) Branching ratio.  The branching ratio $R$, defined as the ratio of
the cross section for capture into the excited state to that for the
ground state, is shown in comparison with the Brune data in
Fig. \ref{fig:tag-branch}.  A weighted least-squares normalization of
the calculation to the laboratory data shows that our calculation of
$R$ is lower than the data by 15\%, which can be combined with the
results above to infer that the excited-state ANC is low by 8\%,
within the range of the Monte Carlo sampling error.  The energy
dependence of the branching ratio matches the data with a $\chi^2$ of
26.0 for 15 degrees of freedom, about as well as the straight line
fits of Brune et al.

3) Energy dependence.  The energy dependence of the \tag\ $S$-factor
at low energy is almost completely independent of cluster-cluster
potential, for the five potentials examined.  After normalizing the
computed cross section (with cluster correlations computed from
potential A of Kim et al. \cite{kim81}) to match the Brune data, we
obtain a $\chi^2$ of 38.7 for 16 degrees of freedom.  A significantly
better fit results if the highest 3 points in energy (where $D$-wave
capture becomes important) are excluded.  The residuals of $-1.8\%$ to
$+8.5\%$ compare favorably with other theoretical calculations, but
the energy dependence of the calculation is systematically shallower
than that of the data at the highest-energy points.  Our calculation
of the $S$ factor gives a logarithmic derivative at $E=0$ of $-0.972\
{\rm MeV}^{-1}$ -- about equal to that found by Mohr et
al. \cite{mohr93}, but half of that found in other theoretical studies
\cite{buck88,kajino86,csoto00}, and about 2/3 of that suggested in a
recent compilation of astrophysical reaction rates \cite{nacre}.

\subsection{\hag}

The low-energy $S$-factor computed from the Kim A cluster-cluster
potential for \hag\ is shown in Fig. \ref{fig:he3ag-breakdown}, along
with the contributions of individual terms of
Eq. (\ref{eqn:multipole}).  Total $S$-factors from five
cluster-cluster potentials are shown along with the laboratory data in
Fig. \ref{fig:he3ag-variation}.  The computed branching ratios are
shown along with the corresponding data in
Fig. \ref{fig:he3ag-branch}.  The discussions of small contributions
to the $S$-factor and of the precision of the results for \tag\ above
also apply here, again with a Monte Carlo error estimate of about 10\%
in $S$-factor normalization.  We again break down the results for
\hag\ into normalization, branching ratio, and energy dependence:

\begin{figure}
\centerline{\epsfig{file=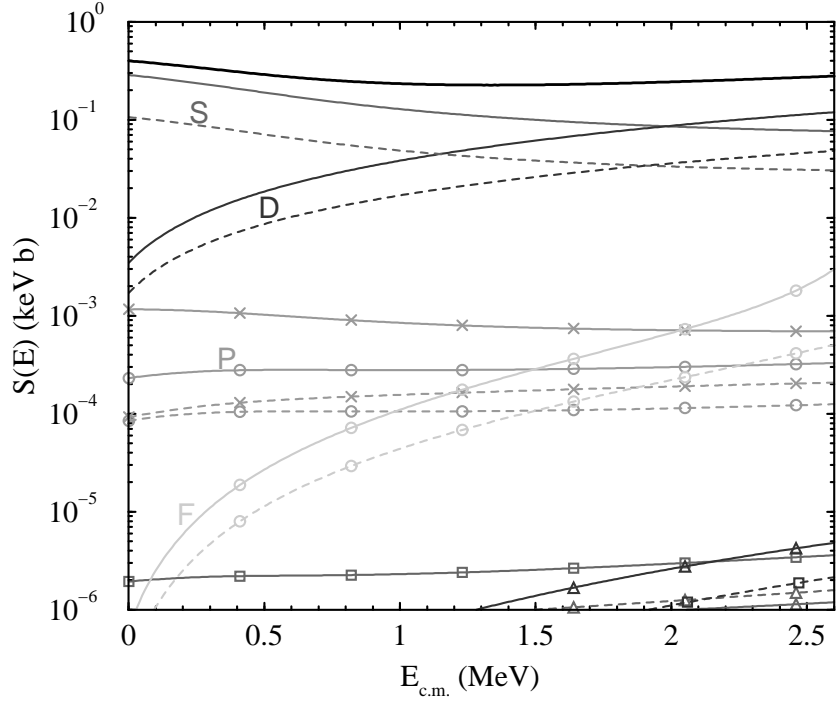,height=11cm,angle=270}}
\caption{Low-energy $S$-factor for \hag\ (thick solid line), and its
breakdown into contributions from various partial strengths, computed
using potential A of Kim et al. \protect\cite{kim81}.  Symbols are as in Fig.
\protect\ref{fig:tag}.}
\label{fig:he3ag-breakdown}
\end{figure}

\begin{figure}
\centerline{\epsfig{file=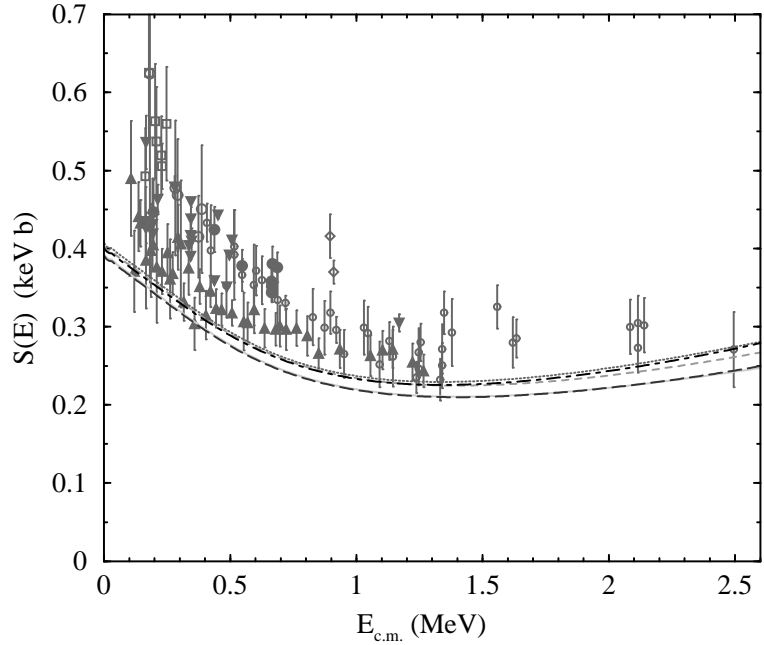,height=10cm,angle=270}}
\caption{Same as Fig. \protect\ref{fig:he3ag-breakdown}, but showing
only total $S$-factors for several different potentials used to
generate cluster correlation functions $\phi_{\alpha t}^{JL}$, with
symbols as in Fig. \protect\ref{fig:tag-variation}.  Data are from
Refs.
\protect\cite{he3ag-hilgemeier,he3ag-kraewinkel,he3ag-nagatani,he3ag-parker,he3ag-robertson,he3ag-osborne}.
Symbols for the data are the same as in Ref.
\protect\cite{adelberger}, with the exception of Refs.
\protect\cite{he3ag-nagatani} ($\Box$) and \protect\cite{he3ag-robertson}
($\diamond$) .}
\label{fig:he3ag-variation}
\end{figure}

1) Normalization.  Our $S$-factors are more than 10\% lower than the
lowest data set (after applying the renormalization of the Kr\"awinkel
data set \cite{he3ag-kraewinkel} recommended by Hilgemeier et
al. \cite{he3ag-hilgemeier}), and nearly a factor of two lower than
the highest data sets (note that this includes the data of Volk et
al. \cite{he3ag-volk}, which are not shown because their results were
published only as extrapolated $S(0)$ values).  By applying five
different phenomenological cluster-cluster potentials that are not in
disagreement with the low-energy elastic-scattering data, we obtain a
small variation in $S(0)$ about a mean of $S(0)=0.40$ keV b.  Using
the potential which best matches the low-energy $S$-wave scattering
(potential A of Kim et al. \cite{kim81}), we obtain $S(0)=0.40$ keV b.
Although this is closer to matching the lower numbers found in capture
photon experiments than the delayed activity experiments, it is not a
close match in normalization to any of the experimental results.  Our
results are therefore not useful for addressing possible systematic
problems in the data.

\begin{figure}
\centerline{\epsfig{file=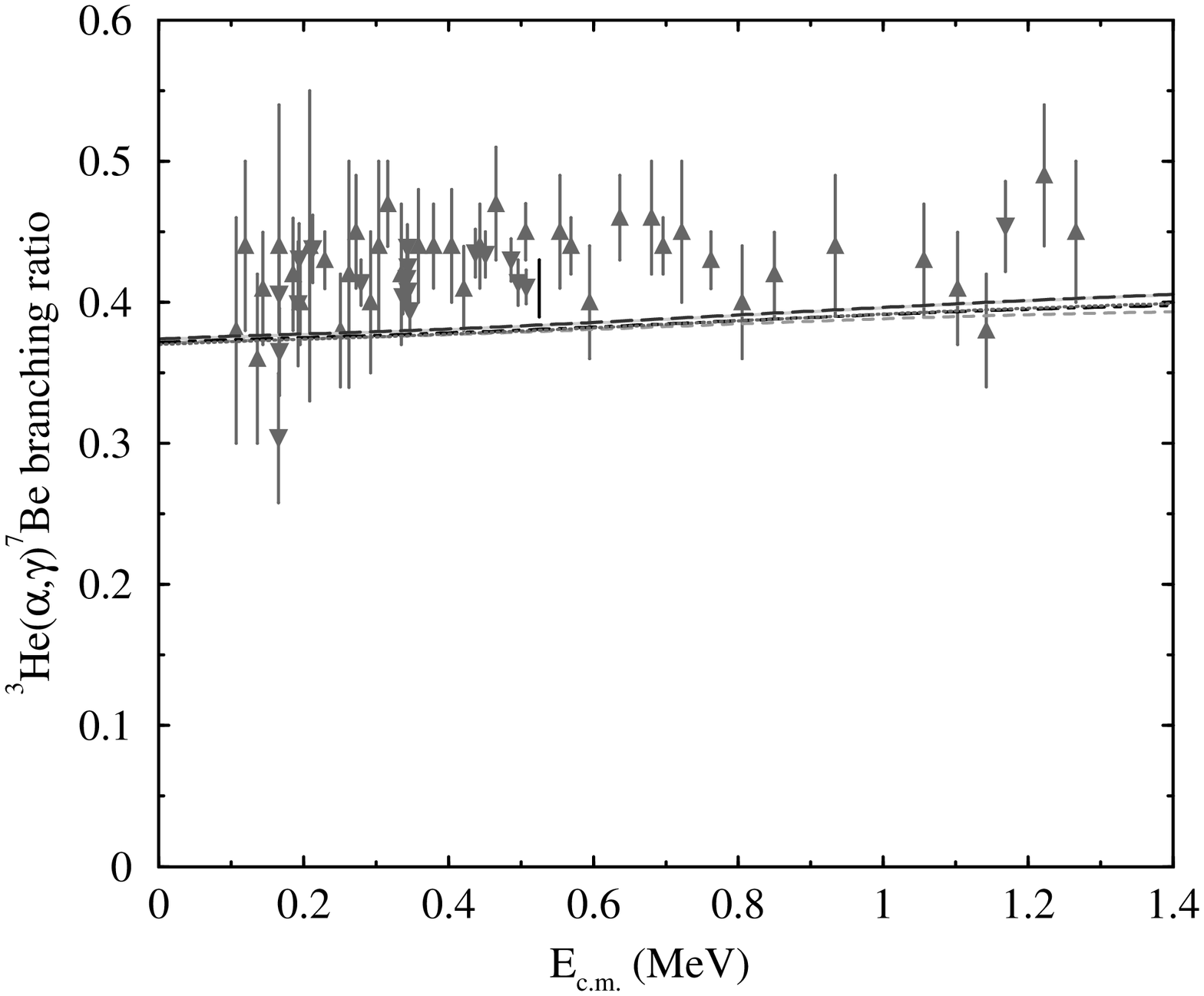,height=10cm,angle=270}}
\caption{Computed ratio of the cross section for capture into the
excited state to that for capture into the ground state, for \hag.
Symbols are the same as in Fig. \protect\ref{fig:he3ag-variation}.}
\label{fig:he3ag-branch}
\end{figure}

2) Branching ratio.  It is seen from the branching ratios in
Fig. \ref{fig:he3ag-branch} that our calculation is in reasonable
agreement with the laboratory data with regard to relative strengths
of transitions to the ground and first excited states of $^7$Be.  This
suggests that the ANCs of the VMC wave functions for both states are
too small, and by about the same factor in each case. Using the
results for $S$-factor normalization above, we conclude that this
factor is in the range 5--25\%.

\begin{table}
\caption{Comparison of energy dependence of calculated $S$-factors
with laboratory data, after renormalizing the computed $S$-factor to
minimize $\chi^2$.  $\nu$ denotes number of degrees of freedom, and an
estimate of systematic normalization uncertainty was subtracted from
the error estimates of the points (as discussed in
Ref. \protect\cite{nollettburles}) before performing this analysis.}
\label{tab:he3ag-energy}
\begin{tabular}{ldr}
Data set & $\chi^2/\nu$ & $\nu$ \\
\hline
Kr\"awinkel \cite{he3ag-kraewinkel} & 0.364 & 37\\
Parker \cite{he3ag-parker}          & 0.917  & 37\\
Hilgemeier \cite{he3ag-hilgemeier}  & 0.698 & 8\\
Nagatani \cite{he3ag-nagatani}      & 0.450 & 6\\
\end{tabular}
\end{table}

3) Energy dependence.  We renormalized our results to best fit each of
the larger data sets separately, and computed chi-squared statistics
in each case to determine goodness of fit.  The results are shown in
Table \ref{tab:he3ag-energy}, and indicate general agreement.  For the
logarithmic derivative of $S(E)$ at $E=0$, we obtain $-0.57\ {\rm
MeV}^{-1}$, in reasonable agreement with other estimates in the
literature \cite{buck88,kajino86,csoto00,nacre}, where published
models fall in the range $-0.50$ to $-0.72$.  It is about equal to the
value presently used in solar neutrino work \cite{adelberger}.

\section{Implications and Applications}
\label{sec:conclusion}

We have carried out the first calculation of the low-energy
$S$-factors for the processes \tag\ and \hag\ based on realistic
nucleon-nucleon potentials.  Seven-body wave functions for these
potentials, constructed by the VMC method and constrained to have the
correct asymptotic forms, produce $S$-factor energy dependences for
the processes \tag\ and \hag\ which agree reasonably well with
experiment.

This work indicates no serious problems in the reaction rates
presently used in astrophysical models.  In fact, the most important
implication of these results for astrophysics is probably that the
previous understanding of these reactions remains essentially
unchallenged.  For example, using the present calculation to
extrapolate the Robertson et al. activity measurement
\cite{he3ag-robertson} of the $S$-factor from 0.9 MeV to 0 MeV, we
obtain essentially the same result as with the energy dependence
currently used in the standard solar model \cite{adelberger}.  For
big-bang nucleosynthesis, there is no low-energy extrapolation
problem.  The most useful result which a theoretical study could
provide for cosmology would therefore be tighter constraints on
cross-section normalizations than the current body of experimental
data provides.  The present calculations have not achieved that goal,
but future first-principles calculations based on realistic
nucleon-nucleon interactions might.

The only serious problem with the results presented here is the low
normalization of the $S$-factors, and its principal cause is probably
easy to identify.  Because most of the cross section arises at large
($> 10$ fm) cluster separations, the low normalizations most likely
arise from the form of the seven-body bound states at large
separations of the p-shell nucleons from the s-shell core.  Part of
the discrepancy may arise from the correlation $f_{sp}$.  The
correlation between $\alpha$ and $\tau$ clusters in the bound states
is proportional to the twelfth power of $f_{sp}$, so that the
long-range correlation between clusters is very sensitive to the
choice of $f_{sp}$.  However, it is hard to see how the $f_{sp}$ used
in this study could affect the wave function beyond about 5 fm cluster
separation.  In the case of \tag, a more likely explanation is that
the Monte Carlo uncertainty has been underestimated, and a
(prohibitively) long integration would ``converge'' on values in
better agreement with experiment.  Other aspects of the present
calculation which are not done as exactly as one may wish, and
therefore may bear on this problem, include the fact that asymptotic
forms have been imposed on the wave functions which are inconsistent
with their energies, as well as the use of cluster-cluster
interactions which were phenomenologically constructed and could
conceivably be inconsistent with the rest of the calculation in subtle
ways.

Despite these lingering difficulties, we have demonstrated the
applicability of a new, almost {\it ab initio} approach for computing
low-energy radiative captures in cases for which precise measurements
exist, achieving the same accuracy as previous theoretical approaches.
We also have presented additional evidence for uncertainty in the
logarithmic derivative of the \tag\ $S$-factor at zero energy.

Regarding the present calculation as a first application of realistic
potentials to radiative capture at mass seven, it has been successful,
and it has taught important lessons about the role of the $f_{sp}$
correlations that were not apparent in previous VMC studies.  This
work clears the way for more refined models of these radiative
captures based on realistic potentials.  Specific improvements which
will be possible in the near future include the use of improved
three-body potentials now in development \cite{PPWC01} and the use of
essentially exact wave functions derived by the Green's function Monte
Carlo technique for the seven-body bound states.  We note that the
application of GFMC will require new wave functions, starting from VMC
wave functions of the type used here to get the asymptotic forms
correct.  In the more distant future, it should be possible to perform
the whole calculation self-consistently, constructing scattering wave
functions from the potentials in a way similar to that used for the
bound-state wave functions.  VMC-based work would also profit by a
more systematic effort to produce a Monte Carlo weighting scheme
well-suited to computing the sorts of matrix elements encounted in
direct capture calculations.  Prospects for significant improvement on
this initial investigation are very good.

\acknowledgments

The author gratefully acknowledges R. Schiavilla, S. C. Pieper,
V. R. Pandharipande, and M. S. Turner for many helpful discussions,
and especially R. B. Wiringa for providing the bound state wave
functions and the variational Monte Carlo code used to compute them,
as well as extensive support in their use.  Computations were
performed on the IBM SP of the Mathematics and Computer Science
Division, Argonne National Laboratory.  This work was supported by the
U. S. Department of Energy, Nuclear Physics Division, under contract
No. W-31-109-ENG-38.  This work was also performed at Argonne National
Laboratory while a Laboratory-Graduate participant, a program
administered by the Argonne Division of Educational Programs with
funding from the U. S. Department of Energy.


\begin{thebibliography}{10}

\bibitem{nollettburles}
K.~M. Nollett and S. Burles, Phys. Rev. D {\bf 62},  123505  (2000).

\bibitem{bntt}
S. Burles, K.~M. Nollett, J.~W. Truran, and M.~S. Turner, Phys. Rev. Lett. {\bf
  82},  4176  (1999).

\bibitem{adelberger}
E.~G. Adelberger {\it et~al.}, Rev. Mod. Phys. {\bf 70},  1265  (1998).

\bibitem{dubovichenko95}
S.~B. Dubovichenko and A.~V. Dzhazairov-Kakhramanov, Phys. Atomic Nucl. {\bf
  58},  579  (1995).

\bibitem{kim81}
B.~T. Kim, T. Izumoto, and K. Nagatani, Phys. Rev. C {\bf 23},  33  (1981).

\bibitem{buck85}
B. Buck, R.~A. Baldock, and J.~A. Rubio, J. Phys. G {\bf 11},  L11  (1985).

\bibitem{buck88}
B. Buck and A.~C. Merchant, J. Phys. G {\bf 14},  L211  (1988).

\bibitem{mohr93}
P. Mohr, H. Abele, R. Zwiebel, G. Staudt, H. Krauss, H. Oberhummer, A. Denker,
  J.~W. Hammer, and G. Wolf, Phys. Rev. C {\bf 48},  1420  (1993).

\bibitem{walliser83}
H. Walliser, Q.~K.~K. Liu, H. Kanada, and Y.~C. Tang, Phys. Rev. C {\bf 28},
  57  (1983).

\bibitem{kajino84}
T. Kajino and A. Arima, Phys. Rev. Lett. {\bf 52},  739  (1984).

\bibitem{walliser84}
H. Walliser, H. Kanada, and Y.~C. Tang, Nucl. Phys. {\bf A419},  133  (1984).

\bibitem{liu81}
Q.~K.~K. Liu, H. Kanada, and Y.~C. Tang, Phys. Rev. C {\bf 23},  645  (1981).

\bibitem{fujiwara83}
Y. Fujiwara and Y.~C. Tang, Phys. Rev. C {\bf 28},  1869  (1983).

\bibitem{kajino86}
T. Kajino, Nucl. Phys. {\bf A460},  559  (1986).

\bibitem{mertelmeier86}
T. Mertelmeier and H.~M. Hoffmann, Nucl. Phys. {\bf A459},  387  (1986).

\bibitem{altmeyer88}
T. Altmeyer, E. Kolbe, T. Warmann, K. Langanke, and H.~J. Assenbaum, Z. Phys. A
  {\bf 330},  277  (1988).

\bibitem{csoto00}
A. Cs\'ot\'o and K. Langanke, Few-Body Systems {\bf 29},  121  (2000).

\bibitem{friedrich78}
H. Friedrich, Nucl. Phys. {\bf A294},  81  (1978).

\bibitem{langanke86}
K. Langanke, Nucl. Phys. A {\bf 457},  351  (1986).

\bibitem{tag-brune}
C.~R. Brune, R.~W. Kavanagh, and C. Rolfs, Phys. Rev. C {\bf 50},  2205
  (1994).

\bibitem{christyduck}
R.~F. Christy and I. Duck, Nucl. Phys. {\bf 24},  89  (1961).

\bibitem{wss95}
R.~B. Wiringa, V.~G.~J. Stoks, and R. Schiavilla, Phys. Rev. C {\bf 51},  38
  (1995).

\bibitem{ppcw95}
B.~S. Pudliner, V.~R. Pandharipande, J. Carlson, and R.~B. Wiringa, Phys. Rev.
  Lett. {\bf 74},  4396  (1995).

\bibitem{a=8}
R.~B. Wiringa, S.~C. Pieper, J. Carlson, and V.~R. Pandharipande, Phys. Rev. C
  {\bf 62},  014001  (2000).

\bibitem{nws}
K.~M. Nollett, R.~B. Wiringa, and R. Schiavilla, Phys. Rev. C {\bf 63},  024003
   (2000).

\bibitem{W91}
R.~B. Wiringa, Phys. Rev. C {\bf 43},  1585  (1991).

\bibitem{ppcpw97}
B.~S. Pudliner, V.~R. Pandharipande, J. Carlson, S.~C. Pieper, and R.~B.
  Wiringa, Phys. Rev. C {\bf 56},  1720  (1997).

\bibitem{APW95}
A. Arriaga, V.~R. Pandharipande, and R.~B. Wiringa, Phys. Rev. C {\bf 52},
  2362  (2000).

\bibitem{WPCP00}
R.~B. Wiringa, S.~C. Pieper, J. Carlson, and V.~R. Pandharipande, Phys. Rev. C
  {\bf 62},  014001  (2000).

\bibitem{WS98}
R.~B. Wiringa and R. Schiavilla, Phys. Rev. Lett. {\bf 81},  4317  (1998).

\bibitem{pieper00}
S.~C. Pieper, private communication.

\bibitem{PPWC01}
S.~C. Pieper, V.~R. Pandharipande, R.~B. Wiringa, and J. Carlson,
  nucl-th/0102004, submitted to Phys. Rev. C.

\bibitem{csoto98}
A. Cs\'ot\'o and K. Langanke, Nucl. Phys. {\bf A636},  240  (1998).

\bibitem{FPPWSA96}
J.~L. Forest, V.~R. Pandharipande, S.~C. Pieper, R.~B. Wiringa, R. Schiavilla,
  and A. Arriaga, Phys. Rev. C {\bf 54},  646  (1996).

\bibitem{brune99}
C.~R. Brune, W.~H. Geist, R.~W. Kavanagh, and K.~D. Veal, Phys. Rev. Lett. {\bf
  83},  4025  (1999).

\bibitem{igamov97}
S.~B. Igamov, R.~M. Tursunmuratov, and R. Yarmukhamedov, Phys. At. Nucl. {\bf
  60},  1126  (1997), [Yad. Fiz. {\bf 60}, 1252 (1997)].

\bibitem{lagaris81}
I.~E. Lagaris and V.~R. Pandharipande, Nucl. Phys. {\bf A359},  349  (1981).

\bibitem{zaikin71}
D.~A. Zaikin, Nucl. Phys. {\bf A170},  584  (1971).

\bibitem{shulgina96}
N.~B. Shul'gina, B.~V. Danilin, V.~D. Efros, J.~M. Bang, J.~S. Vaagen, and
  M.~V. Zhukov, Nucl. Phys. {\bf A597},  197  (1996).

\bibitem{spiger67}
R.~J. Spiger and T.~A. Tombrello, Phys. Rev. {\bf 163},  964  (1967).

\bibitem{ivanovich68}
M. Ivanovich, P.~G. Young, and G.~G. Ohlsen, Nucl. Phys. {\bf A110},  441
  (1968).

\bibitem{boykin72}
W.~R. Boykin, S.~D. Baker, and D.~M. Hardy, Nucl. Phys. {\bf A195},  241
  (1972).

\bibitem{hardy72}
D.~M. Hardy, R.~J. Spiger, S.~D. Baker, Y.~S. Chen, and T.~A. Tombrello, Nucl.
  Phys. {\bf A195},  250  (1972).

\bibitem{enchilada}
J. Carlson and R. Schiavilla, Rev. Mod. Phys. {\bf 70},  743  (1998).

\bibitem{boosts}
J.~L. Forest, V.~R. Pandharipande, and J.~L. Friar, Phys. Rev. C {\bf 52},  568
   (1995).

\bibitem{aps}
A. Arriaga, V.~R. Pandharipande, and R. Schiavilla, Phys. Rev. C {\bf 43},  983
   (1991).

\bibitem{tag-holmgren}
H.~D. Holmgren and R.~L. Johnston, Phys. Rev. {\bf 113},  1556  (1959).

\bibitem{tag-schroeder}
U. Schr\"oder, A. Redder, C. Rolfs, R.~E. Azuma, L. Buchmann, C. Campbell,
  J.~D. King, and T.~R. Donoghue, Phys. Lett. B {\bf 192},  55  (1987).

\bibitem{tag-griffiths}
G.~M. Griffiths, R.~A. Morrow, P.~J. Riley, and J.~B. Warren, Can. J. Phys.
  {\bf 39},  1397  (1961).

\bibitem{tag-burzynski}
S. Burzy\'nski, K. Czerski, A. Marcinkowski, and P. Zupranski, Nucl. Phys. {\bf
  A473},  179  (1987).

\bibitem{tag-utsunomiya}
H. Utsunomiya {\it et~al.}, Phys. Rev. Lett. {\bf 65},  847  (1990).

\bibitem{nacre}
C. Angulo {\it et~al.}, Nucl. Phys. {\bf A656},  3  (1999).

\bibitem{he3ag-hilgemeier}
M. Hilgemeier, H.~W. Becker, C. Rolfs, H.~P. Trautvetter, and J.~W. Hammer, Z.
  Phys. {\bf A329},  243  (1988).

\bibitem{he3ag-kraewinkel}
H. Kr\"awinkel {\it et~al.}, Z. Phys. {\bf A304},  307  (1982).

\bibitem{he3ag-nagatani}
K. Nagatani, M.~R. Dwarakanath, and D. Ashery, Nucl. Phys. {\bf A128},  325
  (1969).

\bibitem{he3ag-parker}
P.~D. Parker and R.~W. Kavanagh, Phys. Rev. {\bf 131},  2578  (1963).

\bibitem{he3ag-robertson}
R.~G.~H. Robertson, P. Dyer, T.~J. Bowles, C.~J. Maggiore, and S.~M. Austin,
  Phys. Rev. C {\bf 27},  11  (1983).

\bibitem{he3ag-osborne}
J.~L. Osborne, C.~A. Barnes, R.~W. Kavanagh, R.~M. Kremer, G.~J. Mathews, J.~L.
  Zyskind, P.~D. Parker, and A.~J. Howard, Nucl. Phys. {\bf A419},  115
  (1984).

\bibitem{he3ag-volk}
H. Volk, H. Kr\"awinkel, R. Santo, and L. Wallek, Z. Phys. {\bf A310},  91
  (1983).

\end{thebibliography}

\end{document}